\documentclass[cameraready]{Interspeech}

\newcommand{\name}{SoniSpeech}
\title{\name: A Large-Scale Open-Vocabulary Tri-Modal Dataset for Wearable Silent Speech Interfaces}

\author[affiliation={1}, orcid=0000-0001-8329-0522, correspondingauthor]{Ruidong}{Zhang}
\author[affiliation={1}, orcid=0009-0007-4525-0352]{Jiacheng}{Liu}
\author[affiliation={1}, orcid=0000-0002-5510-6799]{François}{Guimbretière}
\author[affiliation={1}, orcid=0000-0002-5079-5927, correspondingauthor]{Cheng}{Zhang}

\address{
    $^1$ Cornell University, Ithaca, NY, USA
}

\email{\{rz379, jl4596, fvg3, chengzhang\}@cornell.edu}


\usepackage{comment}

\begin{document}

\maketitle

\begin{abstract}
Wearable silent speech interfaces (SSIs) are limited to small, closed vocabularies. Approaches achieving larger vocabularies require obtrusive hardware such as facial electrodes. We present \name, the first large-scale, open-vocabulary, trimodal dataset for wearable SSI using acoustic-sensing eyewear. It contains 34 hours across 18,000 utterances with three synchronized modalities: ultrasound echo profiles, voiced audio, and frontal video, in both voiced and silent modes. The corpus draws from the SODA dialogue dataset, providing contemporary conversational English with 5,356 unique words and full phoneme coverage. A CTC-based ResNet-34 baseline achieves 26.3\% word error rate (WER) on open-vocabulary silent speech recognition, the first benchmark for this task. Dataset is available at \url{https://doi.org/10.7298/xjjr-9m85}.
\end{abstract}

\begin{figure*}
    \centering
    \includegraphics[width=\linewidth]{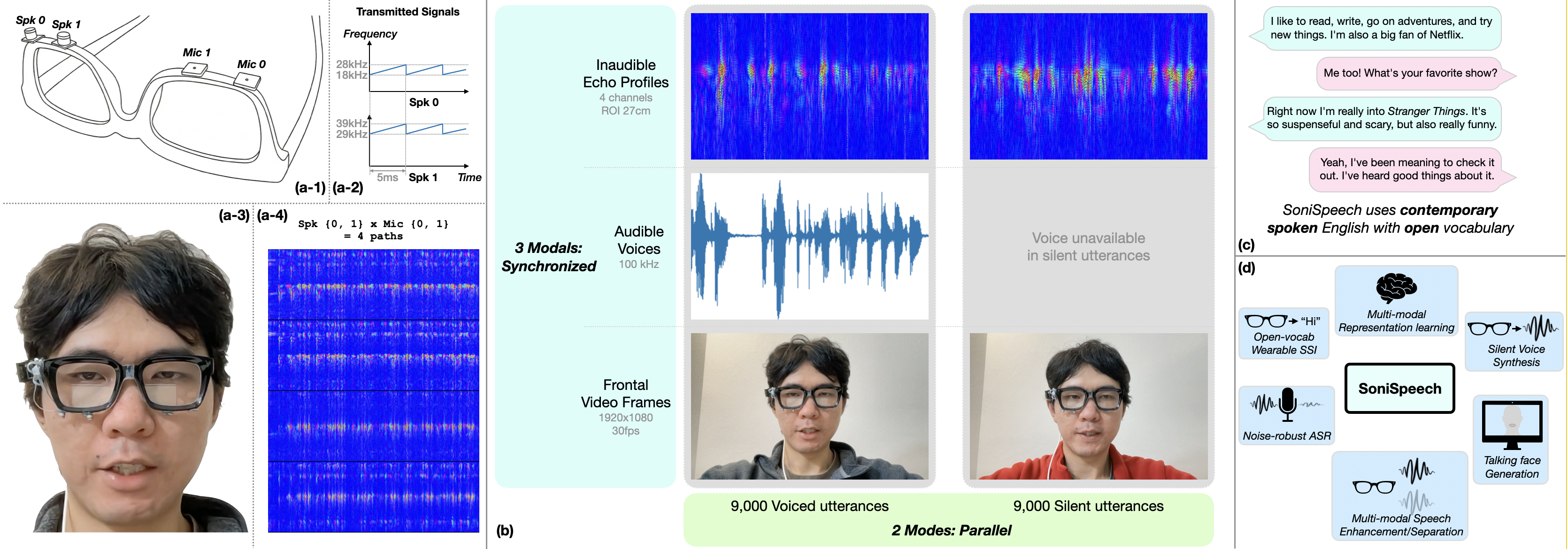}
    \caption{Overview of \name. (a) Hardware and signals. (a-1) Illustration of the hardware setup. (a-2) Configuration of transmitted signals. (a-3) User wearing the \name~device. (a-4) Visualization of 4-channel differential Echo Profiles. (b) Illustration of dataset setup. \name~datasets includes 3 \textit{synchronized} modalities: Echo Profiles as the representation of inaudible signals, audible voices and video frames. The 3 modalities are captured at the same time and are strictly synchronized. Two \textit{paired} modes: voiced and silent utterances were captured. The contents of the utterances are exactly the same. (c) Illustration of the \name~corpus. \name~uses contemporary spoken English corpus with open vocabulary. Samples are drawn from SODA~\cite{kim2023soda}. (d) Illustration of possible research directions enabled by \name.}
    \label{fig:placeholder}
\end{figure*}
\section{Introduction}
Silent speech interfaces (SSIs) offer transformative potential for accessible, private, and low-latency communication~\cite{DENBY2010270}. Yet, practical deployment is currently paralyzed by a fundamental compromise: systems are either capable of open-vocabulary recognition but physically obtrusive~\cite{gaddy2022thesis}, or socially acceptable but confined to small, closed vocabularies~\cite{zhang2023echospeech}. For wearable SSIs to evolve from laboratory prototypes into genuine communication tools, they must break this vocabulary bottleneck and support open-vocabulary, natural language input.

This barrier is not necessarily a limitation of sensing hardware, but rather a critical scarcity of data. Large-scale datasets have proven that open-vocabulary SSI is computationally tractable: facial EMG~\cite{gaddy2022thesis} and ultrasound tongue imaging~\cite{tal2021} corpora have successfully driven down word error rates. However, they demand skin-contacting electrodes or chin-mounted probes. Conversely, minimally-obtrusive wearables such as acoustic-sensing eyewear~\cite{zhang2023echospeech}, depth-sensing devices~\cite{wang2024watchyourmouth} remain trapped evaluating limited isolated commands.

History consistently demonstrates that large-scale, publicly accessible datasets are the ultimate catalyst for paradigm shifts in speech research. Just as LJSpeech~\cite{ljspeech17} and Gaddy's corpus~\cite{gaddy2022thesis} single-handedly unlocked new eras of open-vocabulary research in their respective domains of text-to-speech voice synthesis and EMG-based SSI, unobtrusive wearable SSI requires its own foundational infrastructure to advance. No equivalent resource exists for non-contact, wearable sensors, preventing the community from investigating whether continuous, natural language recognition is even achievable in this form factor.

We bridge this critical gap with \name. By providing the scale and public access necessary to catalyze the next phase of wearable SSI research, we make the following contributions:
\begin{enumerate}
    \item \textbf{The first large-scale, open-vocabulary, trimodal silent speech dataset} collected with minimally-obtrusive acoustic-sensing eyewear: 34 hours, 18,000 utterances, 3 \textit{synchronized} modalities (ultrasound echo profiles, voiced audio, frontal video), in two \textit{parallel} uttering modes: voiced and silent.
    
    \item \textbf{A corpus of contemporary conversational English} constructed from the SODA social dialogue dataset~\cite{kim2023soda}, reflecting natural spoken language rather than historical literary prose.
    
    \item \textbf{The first open-vocabulary baseline} for acoustic-sensing wearable SSI: a CTC-based ResNet-34 system achieving 26.3\% WER on silent speech, demonstrating that the task is tractable and unlocking exciting future directions for wearable SSI.
\end{enumerate}

\section{Related work}

\subsection{Open-vocabulary wearable SSI}
The most successful open-vocabulary SSI efforts rely on high-fidelity but physically obtrusive sensors. Gaddy and Klein~\cite{gaddy2020digital, gaddy2021improved, gaddy2022thesis} created the seminal precedent: approximately 20 hours of 8-channel facial EMG from a single speaker sourced from Project Gutenberg texts. Ultrasound tongue imaging (UTI) offers high-resolution views of the oral cavity. The TaL corpus~\cite{tal2021} provides 24 hours of synchronized UTI, lip video, and audio from 82 speakers, with a small portion being silent speech. The SSR7000~\cite{ssr7000} offers over 7,000 silent utterances with open vocabulary from a single speaker. EEG-based systems have scaled to 175 hours of data from a single participant~\cite{sato2024scaling}. Subsequent advancements have driven WER from 68\% down to below 15\%~\cite{mona2024} for EMG, validating that open-vocabulary SSI is computationally tractable when data is sufficient. However, all of these modalities are confined to laboratory or clinical settings by their bulky and obtrusive form factors: either requiring skin contacting electrodes or chin-mounted probes in mechanically stabilized helmets.

\begin{table}[t]
  \caption{Comparison of datasets. \name~is the first open-vocabulary, trimodal dataset with a minimally-obtrusive commodity form factor. WYM=WatchYourMouth, A=Audio, V=Video, AS=Acoustic Sensing, NS=Not Specified. Sato et al.~\cite{sato2024scaling} is in Japanese.}
  \label{tab:comparison}
  \centering
  \footnotesize
  \begin{tabular}{@{}lccccc@{}}
    \toprule
    \textbf{Dataset} & \textbf{Modality} & \textbf{Vocab} & \textbf{Hours} & \textbf{Spk} & \textbf{Pub}\\
    \midrule
    LJSpeech~\cite{ljspeech17} & A & Open & 24 & 1 & Y\\
    Gaddy~\cite{gaddy2022thesis} & EMG+A & Open & 20 & 1 & Y\\
    TaL~\cite{tal2021} & UTI+A+V & Open & 24 & 82 & Y\\
    Sato et al.~\cite{sato2024scaling} & EEG+A & Open & 175 & 1 & Y \\
    SpeeChin~\cite{zhang2021speechin} & Camera & 54 & - & 12 & NS\\
    EchoSpeech~\cite{zhang2023echospeech} & AS & 31 & - & 12 & NS\\
    Liu et al.~\cite{doi:10.1126/sciadv.ado9576} & IMU & 93 & - & 8 & NS\\
    WYM~\cite{wang2024watchyourmouth} & Depth & 40 & - & 10 & Y\\
    \midrule
    \textbf{\name} & \textbf{AS+A+V} & \textbf{Open} & \textbf{34.1} & \textbf{1} & \textbf{Y}\\
    \bottomrule
  \end{tabular}
\end{table}

\subsection{Minimally-obtrusive wearable SSI}
On the other hand, less obtrusive form factors have witnessed significant improvements yet still confined to small closed vocabulary. Active acoustic sensing~\cite{zhang2023echospeech, zhang2024hpspeech, jin2022earcommand, dong2024rehearsse}, cameras and depth sensing~\cite{zhang2021speechin, wang2024watchyourmouth} and IMUs~\cite{doi:10.1126/sciadv.ado9576, KWON2023105909} offer non-invasive options. However, across all minimally-obtrusive wearable SSI systems, a consistent pattern emerges: they are universally evaluated on small, fabricated command sets. This is not necessarily a limitation of the sensing hardware itself, but rather a consequence of the absence of large-scale training data. For instance, acoustic Echo Profiles, encode millimeter-level facial deformations with high temporal resolution~\cite{zhang2023echospeech, li2022eario, li2024eyeecho}. However, no dataset at the scale needed for open-vocabulary recognition has been collected for any of these form factors.

Open datasets serve as foundational infrastructures for the speech community. Historically, the public release of even single-subject corpora such as LJSpeech~\cite{ljspeech17}, Gaddy's EMG dataset~\cite{gaddy2022thesis}, and recent 175-hour EEG collections~\cite{sato2024scaling} has single-handedly catalyzed rapid, community-wide breakthroughs. Currently, the minimally-obtrusive SSI domain lacks equivalent shared resources at the scale necessary to support open-vocabulary research. Table~\ref{tab:comparison} summarizes how our proposed dataset addresses these critical voids between the proven utility of large-scale public corpora and the specific requirements of unobtrusive wearable SSI.

\section{Dataset description}

\subsection{Sensing hardware}

The sensing platform uses an eyeglass frame equipped with two speakers (Ole Wolff OWR-05049T-38D) and two ultrasound microphones (Syntiant SPH0641LU4H-1) mounted on the lower frame edges. The speakers emit inaudible FMCW chirps that travel across the face; facial deformations during speech alter the signal path, producing detectable echo patterns captured by the microphones. Two FMCW chirp channels are used: 18--28\,kHz and 29--39\,kHz, both with 5\,ms period. Backend electronics consist of a Teensy 4.0 microcontroller with a custom audio board featuring MAX98357A amplifiers and an ADAU7002 PDM-I2S converter, sampling at 100\,kHz. Video is captured from an integrated laptop camera at 1920$\times$1080, 30\,fps. Both inaudible and audible modalities are captured using the microphones on the frame, only separated in frequency in later processing. All three modalities are synchronously captured using a clapping-based synchronization procedure. The platform extends the form factor demonstrated in prior work~\cite{zhang2023echospeech} with a higher sampling rate for expanded bandwidth for more fine-grained movement capture.

\subsection{Corpus construction}

Existing open-vocabulary speech corpora used in SSI research usually draw from formal, written English. LJSpeech~\cite{ljspeech17} uses texts published between 1884 and 1964, Gaddy's corpus~\cite{gaddy2022thesis} draws from 19th-century literary prose. There is a lack of contractions (``I'll'', ``don't''), discourse markers (``you know'', ``like''), colloquialisms (``gonna'', ``y'all''), and contemporary vocabulary (``smartphone'', ``Wi-Fi'') of natural modern speech. Some works use uses fabricated sentence structures~\cite{grid, wang2024watchyourmouth}. A wearable SSI designed for everyday use should be trained on everyday conversational language.

We construct our corpus from SODA~\cite{kim2023soda}, a million-scale social dialogue dataset. SODA provides naturalistic, contemporary conversational English with diverse topics, casual register, contractions, discourse markers, and direct address. These properties closely match real wearable SSI deployment scenarios. Human evaluations have rated SODA dialogues as more consistent, specific, and natural compared to prior dialogue datasets~\cite{kim2023soda}.

We retain only speakable sentences of 5-25 words with ASCII-only characters. URLs, email addresses, non-standard symbols, and dialogue role prefixes are removed. A gender-preserving deterministic name-mapping procedure replaces rare or culturally diverse names with high-frequency US English names to reduce grapheme-to-phoneme out-of-vocabulary rates while keeping proper nouns consistent. Text normalization expands abbreviations and verbalizes numbers (``20'' $\xrightarrow{}$ ``twenty'') and currency symbols (``\$20'' $\xrightarrow{}$ ``twenty dollars''). The final corpus comprises 8,000 training sentences and 1,000 test sentences, sampled from the respective SODA splits to prevent data leakage.

\subsection{Data collection procedure}

Data is collected from a single non-native yet fluent English speaker (an author of this paper) in a controlled quiet environment. The collection spans 360 sessions in two \textit{parallel} modes: 180 voiced (speaking aloud) and 180 silent (mouthing strictly silently without vocalization or whispering), with 160 training and 20 testing sessions per mode. The two modes are \textit{parallel} in that the contents of the utterances are exactly the same. Each session contains 50 utterances. At the start of each session, the speaker claps to produce a synchronization marker aligning wearable device and laptop timestamps. A GUI-based recording interface displays each sentence on screen; the speaker reads aloud or silently, and the system records the utterance text and timestamp before automatically advancing.

The collection procedure prioritizes utterance-level correctness and fluency over recording speed. The speaker re-recorded any utterance with mispronunciations, disfluencies, or hesitations. Breaks were permitted between utterances. Overall, 78.3\% of session time corresponds to valid utterance recordings, with the remainder spent on corrections, re-recordings, and pauses. These idle intervals are removed in post-processing. Post-collection verification identified 5 mismatches between voiced and silent transcripts out of 9,000 utterances (0.06\% mismatch rate), which were corrected in transcripts. 
Due to hardware malfunction, the last 20 sessions of silent modes and last 11 of voiced of training split experienced high noise level on Mic channel 1. However, channels 0 is still available for both audio modals.

\subsection{Dataset statistics}

Table~\ref{tab:stats} summarizes the key dataset statistics. The dataset contains 34.1 hours of session-level data across 18,000 utterances, with 26.7 hours of valid utterance-level recordings. The corpus contains 5,356 unique word types (after normalizing the texts) and 258,702 total word tokens, with 100\% ARPABET phoneme coverage (39/39 phonemes). The test sets contain 1684 unique word types and 242 OOV word types not seen in the training sets.

Silent speech utterances average 5.41\,s compared to 5.26\,s for voiced speech, corresponding to speaking rates of 157 and 162 words/minute respectively, well within the typical range of spoken English~\cite{huang2020speech, barnard2022average}. Silent speech is 2.9\% slower than voiced speech, consistent with the more deliberate articulation observed in silent speech production.

\begin{table}[t]
  \caption{Summary statistics of the \name~dataset.}
  \label{tab:stats}
  \centering
  \footnotesize
  \begin{tabular}{@{}lr@{}}
    \toprule
    \textbf{Statistic} & \textbf{Value} \\
    \midrule
    Total session duration & 34.1\,h \\
    Voiced / Silent sessions & 180 / 180 \\
    Train / Test sentences (per modality) & 8{,}000 / 1{,}000 \\
    Total utterances & 18{,}000 \\
    Avg.\ words per sentence & 14.4 $\pm$ 5.3 \\
    Avg.\ utterance duration (voiced / silent) & 5.26\,s / 5.41\,s \\
    Speaking rate (voiced / silent) & 162 / 157 w/min \\
    Unique word types (total / train / test) & 5{,}356 / 5{,}114 / 1{,}684 \\
    Test OOV word types & 242 \\
    Total word tokens & 258{,}702 \\
    Phoneme coverage & 39/39 ARPABET \\
    Session utilization & 78.3\% \\
    \bottomrule
  \end{tabular}
\end{table}

\subsection{Echo profile feature extraction}

We adopt a similar Echo Profile feature extraction which was demonstrated highly effective in capturing subtle skin deformation~\cite{zhang2023echospeech, zhang2024hpspeech, li2024sonicid, mahmoodi2025echoforce} or larger hand~\cite{yu2024ring, lee2024echowrist, kim2026watchhand, lim2025spellring} and body movements~\cite{mahmud2024actsonic, mahmud2024munchsonic}. Captured audio first goes through a bandpass filter (18-28kHz and 29-39kHz, respectively, corresponding to the transmitted signals), removing any audible components. Each of the 2 speakers $\times$ 2 microphones yields 4 independent FMCW signal paths. For each path, FMCW range processing produces a time-range echo profile. A temporal differentiation removes the static reflections and produces \textit{differential echo profiles}. The differential echo profiles are then cropped to 80 range bins, corresponding to approximately 27.2\,cm of transmission path length. This is sufficient to cover all cross-face paths while minimizing reflections from the environment. The output is a 4-channel differential echo profile sequence at 200\,Hz, used as input to the baseline model.

\section{Baseline system}

\subsection{Overview}

We build a baseline silent speech recognition system to facilitate future research. The baseline encoder is a ResNet-34~\cite{he2016resnet} adapted for sequence modeling. The input is a 4-channel differential echo profile with 80 range bins along the frequency axis. The standard stem max-pool layer is removed to preserve temporal resolution for phonetic recognition; temporal downsampling is instead performed via strided convolutions in the stem and residual stages 2, 3, and 4, yielding a total 16$\times$ temporal downsampling factor. The 2D feature maps are collapsed to a 1D sequence via frequency-wise global average pooling, producing a 512-dimensional embedding at each time step. A linear projection maps these embeddings to output logits. Group Normalization (32 groups) is used throughout for stable convergence with small batch sizes.

A SentencePiece~\cite{kudo2018sentencepiece} Unigram tokenizer with a vocabulary of 1{,}000 units is trained on the transcript corpus. This setup is close to practical ASR deployment and supports generalization to unseen words.

\subsection{Training details}

Models are trained with CTC loss~\cite{graves2006ctc} using Adam (learning rate $2 \times 10^{-4}$, weight decay $10^{-4}$) and Exponential Decay scheduler with linear warm-up, for 200 epochs with batch size 16. Data augmentation consists of SpecAugment~\cite{park2019specaugment} (2 frequency masks, 3 time masks) and concatenation of up to 3 utterances or 15\,s per sample for extended context modeling.


Models are trained on three data configurations, on the first 140 sessions from both modes: \textit{voiced-only}, \textit{silent-only}, and \textit{voiced + silent combined}. This setup directly tests whether voiced speech data can supplement silent speech training and whether the two modalities provide complementary or conflicting training signals.

\begin{figure}
    \centering
    \includegraphics[width=\linewidth]{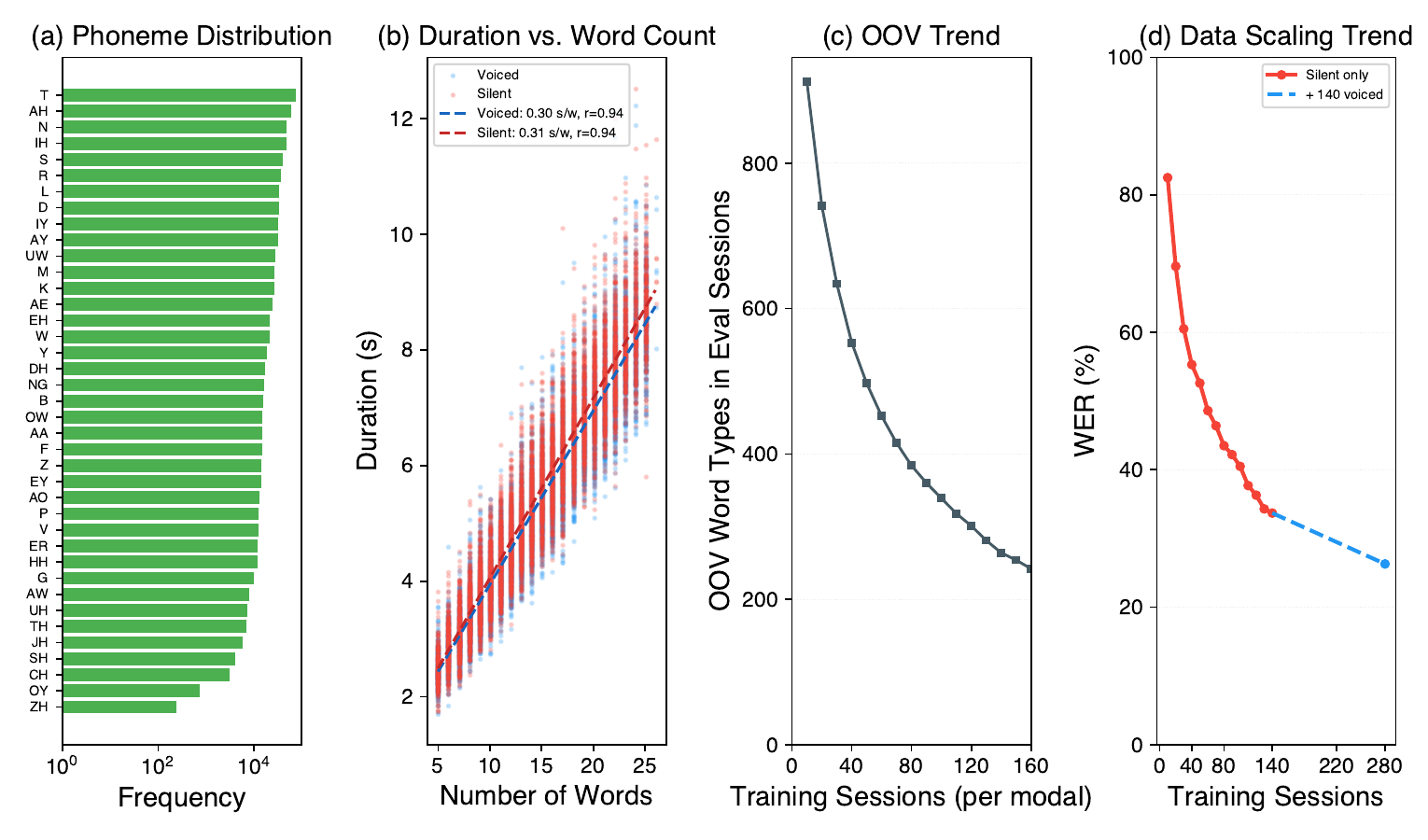}
    \caption{Statistics and results. (a) Distribution of ARPAbet phoneme frequency in the \name~corpus. (b) Distribution of utterance duration vs. number of words. (c) Number of OOV word types in the evaluation set as the number of training sessions scales. (d) Performance curve as number of training sessions scales.}
    \label{fig:results}
\end{figure}

\section{Experiments and results}

\subsection{Main results}

Table~\ref{tab:results} summarizes the performance for each configuration.

\begin{table}[t]
  \caption{Summary of performance.}
  \label{tab:results}
  \centering
  \footnotesize
  \begin{tabular}{@{}lcc@{}}
    \toprule
    & \multicolumn{1}{c}{\textbf{Eval: Voiced}} & \multicolumn{1}{c}{\textbf{Eval: Silent}} \\
    \cmidrule(lr){2-2} \cmidrule(lr){3-3}
    \textbf{Training Data} & WER & WER \\
    \midrule
    Voiced only    & 16.9\% & 78.4\% \\
    Silent only    & 55.7\% & 33.7\% \\
    Voiced + Silent & 15.8\% & 26.3\% \\
    \bottomrule
  \end{tabular}
\end{table}

\textbf{Silent-to-silent recognition.}
The silent-only model achieves 33.7\% WER on open-vocabulary silent speech recognition. To our knowledge, this is the first time open-vocabulary WER has been reported for an unobtrusive acoustic-sensing SSI. At this level, the system produces meaningful output that captures the structure and content of the spoken sentences, demonstrating that the task is tractable. This is a baseline result using a standard ResNet encoder, CTC loss, greedy decoding, and no external language model, leaving substantial room for improvement.

\textbf{Voiced-to-voiced validation.}
The strong voiced-speech performance (16.9\% WER) confirms that ultrasound echo profiles contain sufficient articulatory information for speech recognition, even without any audible components. This points a promising direction for multi-modal speech recognition, voice enhancements, voice separation and related fields.

\textbf{Cross-modal mismatch.}
A large gap separates voiced-trained and silent-evaluated models ($>$50\% WER) from within-modality performance. This mismatch is a scientific finding, not just a failure mode. Three factors contribute: (1) different articulation dynamics: silent speech is slower (2.62 vs.\ 2.70 words/second in our data) with more deliberate mouth movements and different co-articulation patterns; (2) voiced acoustic energy extending into the ultrasonic band (18--39\,kHz), where plosive and fricative consonants directly affect echo profiles, providing discriminative features absent in silent speech; (3) differing muscle engagement patterns producing different co-articulation transitions between phonemes, which was also confirmed in previous research~\cite{gaddy2022thesis}. Pre-training on voiced speech alone is therefore insufficient for silent speech recognition, and future work must explicitly address this domain gap.

\textbf{Complementary training signal.}
The combined voiced+silent model achieves significantly improved (from 33.7\% to 26.3\%) performance than silent-only training. This confirms that voiced data provides a complementary training signal. This can inspire future multi-task and contrastive learning approaches. Considering that this baseline performance does not use any language model, foundation model training or contrastive learning, future room for improvement is massive.

\subsection{Data scaling}
To examine the effectiveness of scaling training data, we iterate the amount of silent-only training data from 10 to 140 sessions (the last 20 sessions with only 1 working mic channel were not used to ensure all training sessions are equivalent) and evaluate the corresponding performance, visualized in Figure~\ref{fig:results}(d). A steady improvement in performance is observed as training data scales. At 140 sessions, it has not saturated. Adding additional 140 voiced sessions further improves performance, even though the modality is different. This confirms the value of a large-scale dataset and encourages future contribution.
\section{Discussion and conclusion}

We have introduced \name, the first large-scale, open-vocabulary, trimodal dataset for minimally-obtrusive acoustic-sensing silent speech recognition. The baseline WER of 26.3\% on silent speech demonstrates that open-vocabulary wearable SSI is a tractable problem, while the significant voiced-silent modality gap reveals a critical domain adaptation challenge for future work.

By design, our data collection focuses on a single speaker to isolate the vocabulary-scale challenge before tackling inter-speaker variability. This follows the approach of LJSpeech~\cite{ljspeech17}, Gaddy's EMG corpus~\cite{gaddy2022thesis}, SSR7000~\cite{ssr7000} and Sato et al.\cite{sato2024scaling}, which are all single-speaker datasets that catalyzed their respective research directions. The acoustic-sensing eyewear platform has already demonstrated multi-user robustness~\cite{zhang2023echospeech}, indicating that multi-speaker extension is feasible. Our data collection is conducted in a single quiet environment. However, EchoSpeech~\cite{zhang2023echospeech} has demonstrated robustness to walking and noisy conditions, suggesting the sensing approach generalizes. The baseline intentionally uses no external language model or pre-trained representations, establishing a clean lower bound that the community can build upon.

The trimodal alignment of \name~opens several future research directions in addition to silent speech recognition: cross-modal contrastive learning between video, voiced audio and silent ultrasound streams following approaches such as MONA~\cite{mona2024}; self-supervised pre-training from paired video and audio; speech synthesis from silent input~\cite{gaddy2020digital}; extension to multiple speakers and beyond. The dataset is released to accelerate community research on this underexplored problem.

\section{Use of Generative AI Disclosure}
Generative AI has been used to improve writing of this paper. The models used include: Gemini 3.0 Pro and Claude Opus 4.6.

\section{Acknowledgments}
This research is supported by the National Science Foundation Grant No. 2239569. The lead author is partially supported by Qualcomm Innovation Fellowship. We would also like to thank the Information Science Department at Cornell University for providing support for this work.



\bibliographystyle{IEEEtran}
\bibliography{mybib}

\end{document}